\documentclass[10pt,aps,prb,twocolumn,superscriptaddress,floatfix]{revtex4-2}
\usepackage[margin= 0.6in,top= 18.0mm]{geometry}
\usepackage{amsmath,amssymb,bm} 
\usepackage{graphicx}
\usepackage{epstopdf}
\usepackage{latexsym}
\usepackage[dvipsnames]{xcolor}
\usepackage{braket}
\usepackage{float}
\usepackage[normalem]{ulem}
\usepackage{comment}
\usepackage{mathtools}
\usepackage{array}
\usepackage{tabu}
\usepackage{multirow}
\usepackage[inline]{enumitem}
\usepackage[normalem]{ulem}
\usepackage{natbib}  
\usepackage[version=4]{mhchem}
\usepackage{csquotes}
\usepackage[colorlinks=true,linkcolor=blue,citecolor=blue]{hyperref}
\usepackage{cleveref}

\begin{document}
\title{Evolution of inhomogeneities in two-dimensional disordered superconductors in a magnetic field}
\author{Poulami Sarkar}
\affiliation{Department of Physical Sciences, Indian Institute of Science Education and Research Kolkata, Mohanpur 741246, India}
\author{Jhinhwan Lee}
%\thanks{Email: jhinhhwan@ibs.re.kr}
\affiliation{Center for Artificial Low-Dimensional Electronic Systems, Institute for Basic Science (IBS), Pohang 37673, Republic of Korea}
\author{Hae Ryoung Park}
\affiliation{Center for Artificial Low-Dimensional Electronic Systems, Institute for Basic Science (IBS), Pohang 37673, Republic of Korea}
% --- New additional affiliation for H.-R. Park as in AdMa'24 ---
\affiliation{Department of Physics, Pohang University of Science and Technology (POSTECH), Pohang 37673, Republic of Korea}
\author{Anushree Datta}
\affiliation{Laboratoire Matériaux et Phénomènes Quantiques, Université de Paris, CNRS, F-75013 Paris, France}
\affiliation{Laboratoire de Physique des Solides, Université Paris-Saclay, CNRS, 91405 Orsay, France}
\affiliation{Department of Applied Physics, Aalto University School of Science, FI-00076 Aalto, Finland}
\author{Amit Ghosal}
\affiliation{Department of Physical Sciences, Indian Institute of Science Education and Research Kolkata, Mohanpur 741246, India}

\begin{abstract}
Emerging granularity in superconducting films by tuning disorder is a well-studied topic, both theoretically and experimentally. However, the orbital magnetic field generates a vortex lattice and contributes to the formation of periodic inhomogeneities. Here, we study superconducting films in the simultaneous presence of disorder and a magnetic field, examining how inhomogeneities in various superconducting correlations evolve under these two perturbations. By performing scanning tunneling spectroscopy (STS) on thin films of \ce{Sr2VO_{3-\text{x}}FeAs} layer structures under both zero and finite orbital magnetic fields, we report impressive similarities between our theoretical results and the experimental findings. Our results have strong implications for identifying the nature of vortices in disordered superconductors, demonstrating a crossover from Abrikosov to Josephson character with increasing disorder, and provide predictive guidance for interpreting STS and current mapping data in complex superconductors.
\end{abstract}
\maketitle

\section{\label{sec:level1}INTRODUCTION}
The individual response of disorder and an orbital magnetic field, $H$, on an s-wave superconductor (SC) has been studied extensively~\cite{dobrosavljevic2012conductor, Sherman:2014dma, RevModPhys.76.975}. However, the interplay of an SC with simultaneous perturbation of disorder and magnetic field gives rise to rich and unconventional phenomena, which are crucial for understanding smart quantum materials. While disorder disrupts SC coherence by introducing spatial fluctuations, a magnetic field generates vortices that modulate the SC state. When both perturbations act simultaneously, their competition leads to new emergent phases and unconventional transitions that differ significantly from their individual effects~\cite{PhysRevB.107.L140502}. Investigating this combined effect is essential for fundamental physics and technological applications in SC devices~\cite{PhysRevB.67.184515,tinkham2004introduction, RevModPhys.66.1125}, where disorder and external fields are inevitable.

Disorder, particularly in the form of non-magnetic impurities, induces spatial variations in the SC pairing amplitude, $\Delta$, creating SC islands of size $\xi$ (coherence length) separated by insulating regions~\cite{PhysRevB.65.014501, PhysRevLett.81.3940, Bouadim2010SingleAT, article1}. As disorder increases, superconductivity weakens due to the fragmentation of coherent pairing, ultimately driving a transition from a SC state to an insulating state, driven by the depletion of the single-particle energy gap ($E_g$)~\cite{doi:10.1126/science.1149587,sacepeb, PhysRevB.75.184530} and the loss of superfluid phase stiffness ($D_s$)~\cite{PhysRevB.71.014514, PMID:17943125, PhysRevLett.106.047001}, both of which are crucial for maintaining global phase coherence.

In contrast, a magnetic field perturbs superconductivity by introducing Abrikosov vortices~\cite{Abrikosov:1957wnz}, each featuring a normal metallic core of size $\xi$ and circulating supercurrents that extend over the penetration depth $\lambda$. Unlike disorder, which disrupts superconductivity by localizing Cooper pairs, the field-driven transition occurs through the proliferation of vortices. As the vortex density increases with the applied field, their cores begin to overlap, disrupting the SC state and leading to a transition from a SC to a metallic phase at a critical field $H_c$~\footnote{In this paper, we considered strongly type-II SC as the first critical field for the insertion of a single vortex $H_{c1} \rightarrow 0$. For simplicity in notation, $H_c$ commonly denotes $H_{c2}$ in the standard literature.}.

When disorder and a magnetic field act together, the nature of superconductivity is altered in ways not seen in either case alone~\cite{PhysRevB.107.L140502} -- (a) The critical field $H_c$ at which superconductivity is lost decouples from the critical field for the vanishing of the spectral gap as the disorder strength, $V$, increases~\cite{article_indranil}. (b) Strong disorder alters the nature of SC vortices -- the Abrikosov vortices realized at weak disorder gradually lose their metallic cores, transforming into coreless Josephson vortices~\cite{PhysRevB.49.6416, Radovan2006-RADAVL, RevModPhys.91.011002} at strong disorder. These change the fundamental nature of the normal state obtained by field-induced transition in weak and strong disorder, and yield a complex phase diagram in the $V-H$ plane at zero temperature.

In this article, we explore how inhomogeneities in local observables that characterize superconductivity, such as the local SC pairing amplitude, local coherence peak height, local density of states (LDoS), and local SC currents, evolve with disorder strength, $V$, and the orbital field strength, $H$. A key focus of our study is to investigate any spatial correlations in these observables. Experimental studies, particularly those utilizing scanning tunneling spectroscopy (STS), address many of these observables~\cite{PhysRevLett.85.1536, RevModPhys.79.353}, and in particular have uncovered distinct features in the LDoS that differentiate Abrikosov and Josephson vortices in disordered SC~\cite{article}. By comparing our results with experimental observations, we establish the efficiency of our theoretical calculations, which make predictions about other SC correlations that are hard to track in experiments.

The rest of the paper is organized as follows: In Sec. \ref{sec:level2}, we describe the model and parameters, as well as the key points of our simulation method. We also describe our experimental setup. In Sec. \ref{sec:level3}, we present our results on the inhomogeneities in various SC correlations, such as SC pairing amplitude, coherence peak-heights, local density of states, and current patterns. We emphasize the similarities in our theoretical predictions and experimental observations. We also include comprehensive discussions and implications of our findings. Finally, we conclude in Sec. \ref{sec:level4}.
%%%%%%%%%%%%%%%%%%%%%%%%%%%%%%%%%%%%%%%%%%%%%%%%%%%%%%%%%%%%%%%
\section{\label{sec:level2}Model and Method}
We describe a type-II SC film by the 2D attractive Hubbard model with non-magnetic impurities and orbital magnetic field:
\begin{equation}
    \begin{split}
        & \hat{\mathcal{H}}=\hat{\mathcal{H}}_0+\hat{\mathcal{H}}_{int} \\
        & \hat{\mathcal{H}}_0 =-\sum\limits_{\langle i,j \rangle ,\sigma}(te^{-i\phi_{ij}}\hat{c}^\dagger_{i\sigma}\hat{c}_{j\sigma} + H.c. ) + \sum\limits_{i\sigma}(V_i-\mu)\hat{n}_{i\sigma} \\
        & \hat{\mathcal{H}}_{int}=- |U|\sum\limits_i\hat{n}_{i\uparrow}\hat{n}_{i\downarrow}
    \end{split}
\end{equation}
Here, the first term represents the tight-binding model, which describes the hopping of spin-$1/2$ electrons on a 2D square lattice, where $t$ is the hopping parameter. $\hat{c}^\dagger_{i\sigma} (\hat{c}_{i\sigma})$ are electronic creation (annihilation) operators, and $\hat{n}_{i\sigma}=\hat{c}^\dagger_{i\sigma}\hat{c}_{i\sigma}$ is the spin-resolved electronic density operator at site $i$ with spin $\sigma$. We use $\hat{n}_i=\sum_{\sigma}\hat{n}_{i\sigma}$ to represent the spin-summed local density operator, and the average density of charge carrier is then represented by $\rho=\langle \hat{n}_i \rangle$. We also assume no magnetization in the system, even locally, so $\rho_{i\uparrow}=\rho_{i\downarrow}$. The orbital magnetic field is derived from a vector potential, $\textbf{H}=\nabla \times \textbf{A}$ and the vector potential $\textbf{A}$ is introduced via the Peierls phase factor~\cite{rpeierls}, $\phi_{ij}=\frac{\pi}{\phi_0}\int_i^j\textbf{A}\cdot d\textbf{l}$, where $\phi_0=\frac{h}{2e}$ represents the SC flux quantum. We consider a uniform magnetic field $\textbf{H}=H\hat{z}$ and choose the Landau gauge $\textbf{A}=(0,-Hx)$. We model the disorder as (non-magnetic) impurities with site energies $V_i$, which are uniformly distributed within the range [$-V, V$]. Thus, $V$ represents the strength of disorder in our model. The chemical potential $\mu$ fixes the average electronic density $\rho = \sum_{i\sigma} \rho_{i\sigma}$ in the system. We measure all energies in units of $t$ and all distances in units of lattice spacing $a$.

In our calculation, we evaluate the local SC (complex) order parameter
$\Delta_i=-|U|\langle c_{i\uparrow}c_{i\downarrow}\rangle$ and the local density $\rho_i=\langle c^\dagger_{i\uparrow}c_{i\downarrow}\rangle$ for all sites $i$ for a given $U$, $\rho$ $V$ and $H$  using iterative self-consistency of the Bogoliubov-de–de Gennes (BdG) technique. The BdG calculations are performed on a magnetic unit cell (MUC) of typical size $50 \times 50$, and the vortex lattice is formed in the absence of disorder by periodically repeating~\cite{PhysRevB.66.214502} the MUCs forming the supercell across which the periodic boundary condition is applied. Each MUC contains an even number of SC flux quanta $\phi_0$. In particular, we will discuss below the results for two strengths of magnetic fields -- $\phi/\phi_0=2,6$. Thus, our vortex lattice, composed of $5 \times 5$ MUCs, contains a total of $50$ and $150$ SC-vortices. Upon achieving the self-consistency in $\Delta_i$ and $\rho_i$ for all sites $i$ and in $\mu$, we calculate different SC correlation functions using the final BdG eigenvectors and eigenvalues. Results for all observables are averaged over up to $10$ different realizations of disorder configurations. While we studied our model for a range of parameters in ${\mathcal{H}}$ of Eq.~(1), we focus on the results for parameters $U=1.2$ and $\rho=0.875$ in the following.

% ===== New experimental paragraph (entirely new → fully underlined) =====
Scanning tunneling measurements on \ce{Sr2VO_{3-\text{x}}FeAs} followed Ref.~\cite{article}. Differential conductance was obtained using a numerical $dI/dV$ method. Unless stated otherwise, maps covered $270\,\mathrm{nm}\times270\,\mathrm{nm}$ with pixel size $\approx 0.586\,\mathrm{nm}$ at $\mu_0H=0$ and $7$~T. For quantitative comparison to theory we use analysis windows closely matched to the theory box ($50\times50$ sites $\approx 13.8\,\mathrm{nm}\times13.8\,\mathrm{nm}$, taking Fe–Fe spacing $a\approx 0.277\,\mathrm{nm}$) centered at regions of large (high effective disorder) and small (low disorder) zero-bias LDoS. Where helpful, we quote fields both as reduced flux $\phi/\phi_0$ and as $\mu_0H$ via the chosen MUC.
 %%%%%%%%%%%%%%%%%%%%%%%%%%%%%%%%%%%%%%%%%%%%%%%%%%%%%%%%%%%%%

\section{\label{sec:level3}RESULT}

We present our results below, discussing the nature and degree of inhomogeneity in the system in response to the combined effects of $V$ and $H$.

\subsection{\label{subsec:inhom}Nature and degree of inhomogeneity}
%%%%__________________________________________________________
\begin{figure}[t]
    \centering
    \includegraphics[width=0.48\textwidth,height=0.45\linewidth]{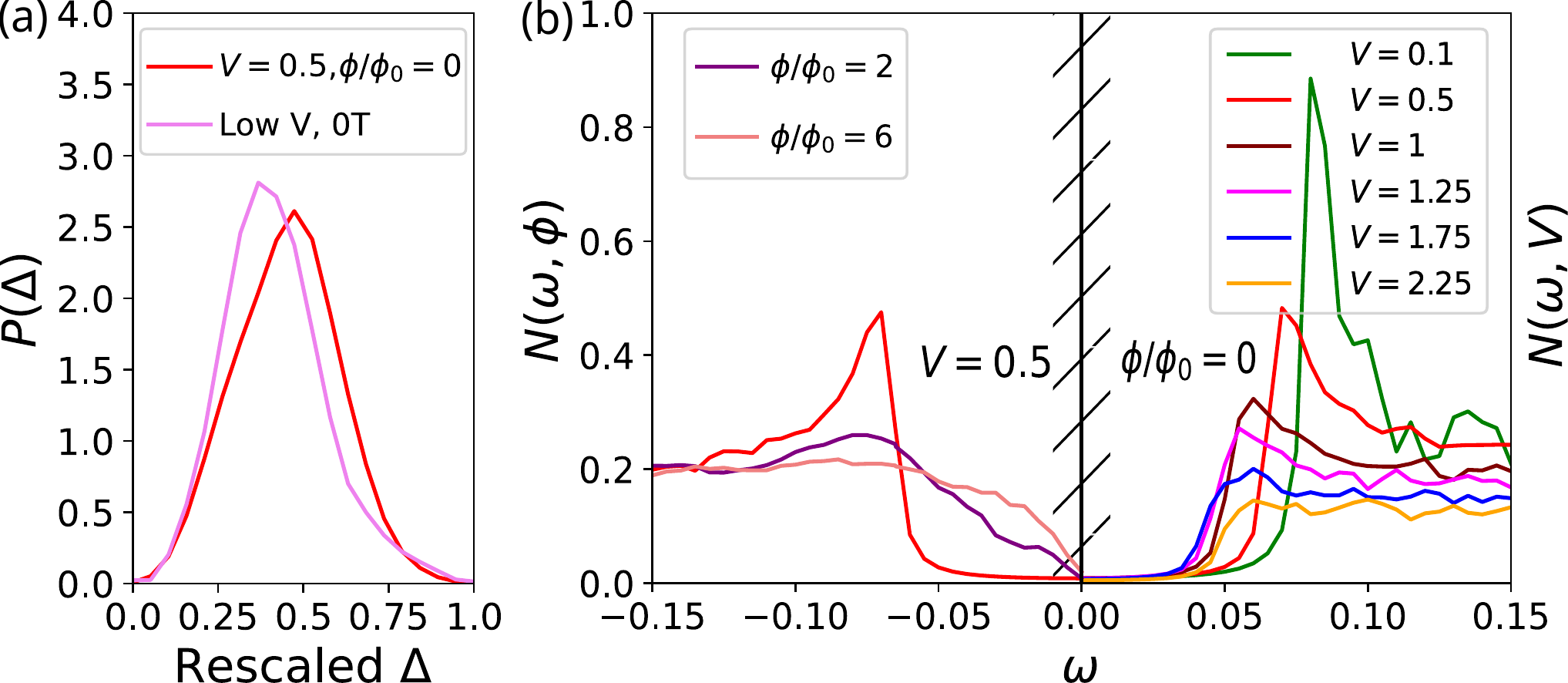}
    \caption{Panel (a) shows normalized
distributions of SC pairing amplitude $P(\tilde{\Delta})$: The red trace is for $\tilde{\Delta}^{\rm theo}$ simulated using the BdG method on a $50 \times 50$ system for $V=0.5$ in the absence of a magnetic field. The violet trace represents $P(\tilde{\Delta}^{\rm expt})$ obtained from the experiments with Low disorder at $5\,\text{K}$ in zero field. $P(\tilde{\Delta}^{\rm theo})$ has a standard deviation, $\sigma_{\tilde{\Delta}^{\rm theo}} = 0.1534$, whereas $P(\tilde{\Delta}^{\rm expt})$ yields a similar value for the standard deviation, $\sigma_{\tilde{\Delta}^{\rm expt}} = 0.1518$, confirming that the extent of disorder in these two complementary methods are broadly similar. Panel (b) displays the evolution of the {\it average} density of states (AvDoS), $N(\omega)$, of the 2D SC (for system size $50 \times 50$). Exploiting the particle–hole symmetry of the SC state, we plot only one side of $\omega$ for the spectrum. The left part of the panel shows $N(\omega)$ for fixed disorder strength $V=0.5t$ at different magnetic fields $\phi/\phi_0 = 0, 2, 6$ over the range $\omega \in [-0.15, 0]$. In contrast, the right part presents $N(\omega)$ for different $V$ at zero field ($\phi/\phi_0=0$) over the range $\omega \in [0, 0.15]$.}
 
    \label{inhom-avdos}
\end{figure}
%%%%__________________________________________________________
We begin our results by examining the probability distribution $P(\Delta)$ of the SC pairing amplitude, which counts the frequency of the occurrence of the magnitude of the SC pairing amplitude somewhere in the system, as a function of $\Delta$ itself. $P(\Delta)$ provides a measure of the spatial inhomogeneity in the system. To ensure a justified comparison between our simulations and experiments, we present $P(\Delta)$ from our simulation and experiment in Fig.~\ref{inhom-avdos}(a)~\cite{PhysRevB.96.054509}. Here, the distributions are plotted for rescaled variables $\tilde{\Delta}_i = (\Delta_i - \Delta_{\min})/(\Delta_{\max} - \Delta_{\min})$, which maps corresponding SC pairing amplitudes onto a universal normalized scale $\in [0,1]$. We notice that $P(\tilde{\Delta})$ from both the simulation and experimental has a near-normal form with similar width, confirming a fair comparison between the inhomogeneities in the two.

Having established a ground for fair comparison between our theory and experiments, we turn to the spectral features to capture how inhomogeneity evolves under the influence of disorder $V$ and the magnetic field $\phi/\phi_0$. A pristine SC exhibits a hard gap, $E_g^0=\Delta_0$, in the single-particle excitation spectrum, which is reflected in the single-particle density of states (DoS) $N(\omega)$. $E_g^0$ is identical to the uniform SC pairing amplitude $\Delta_0$ of the BCS calculation. The low-energy states around the Fermi energy of the underlying metal are pushed to the gap edge $E_g^0$, producing sharp coherence peak heights (CPH) whose height diverges in the thermodynamic limit. The introduction of $V$ (keeping $\phi/\phi_0=0$) reduces $E_g^V$ (which is the $V$-dependent energy gap) and depletes CPH progressively until a critical $V_c$, by which the CPH disappears in the background DoS, and above $V_c$, $E_g$ keeps increasing~\cite{PhysRevB.71.014514, PhysRevLett.81.3940}. Here, $\phi_0$ is the flux through a $50 \times 50$ system containing one SC vortex. This $V_c$ where $E_g$ is at a minimum is a good measure for the occurrence of superconductor-insulator transition driven by the disorder~\cite{PhysRevB.71.014514, PhysRevLett.81.3940}. Such a behavior of $N(\omega)$ is illustrated in Fig.~\ref{inhom-avdos}(b)[right].
The development of $N(\omega)$ with finite $\phi$ for a weakly disordered ($V=0.5$) SC is different, as we illustrate in Fig.~\ref{inhom-avdos}(b)[left].
In addition to CPH falling rapidly with increasing $\phi$, significant low-lying states are induced by $H$, though the gap-filling remains partial. A near V-shaped soft gap in the AvDoS persists until $\phi/\phi_0 = 6$. In particular $N(\omega=0)\approx 0$. This soft gap, however, fills up completely for stronger magnetic fields~\cite{PhysRevB.107.L140502}.

Interestingly, the trends depicted in the Fig.~\ref{inhom-avdos}(b)[right] help us define local coherence peak height (LCPH) in the following manner:
%%%%%%%%%%%%%%%%%%%%%%%%%%%%%%
%Here, we have formulated the local coherence peak height at different disorder strengths to be 
\begin{equation}
     h_i=\frac{1}{2}\left[N(i,\omega=E_g^V)+N(i,\omega=-E_g^V)\right]-C_V
\end{equation}
where $C$ is a $V$-dependent constant- the value of the DoS of the underlying normal state at those SC gap edge energies averaged between positive and negative bias, given by,
\begin{equation}
     C_V=\frac{1}{2}\left[N_{\rm Norm}(i,\omega=E_g^V)+N_{\rm Norm}(i,\omega=-E_g^V)\right]
\end{equation}

Such construction ensures that $h_i \in [h^0,0]$, and the $\langle h_i \rangle$ decreases monotonically from $h^0$ to $0$, as $V$ and/or $\phi$ is tuned from zero to large values.
Here, $N(i,\omega)$ is the LDoS at site $i$ for energy $\omega$.
%%%%__________________________________________________________
\begin{figure}[ht]
    \centering
    \includegraphics[width=0.48\textwidth,height=1.5\linewidth]{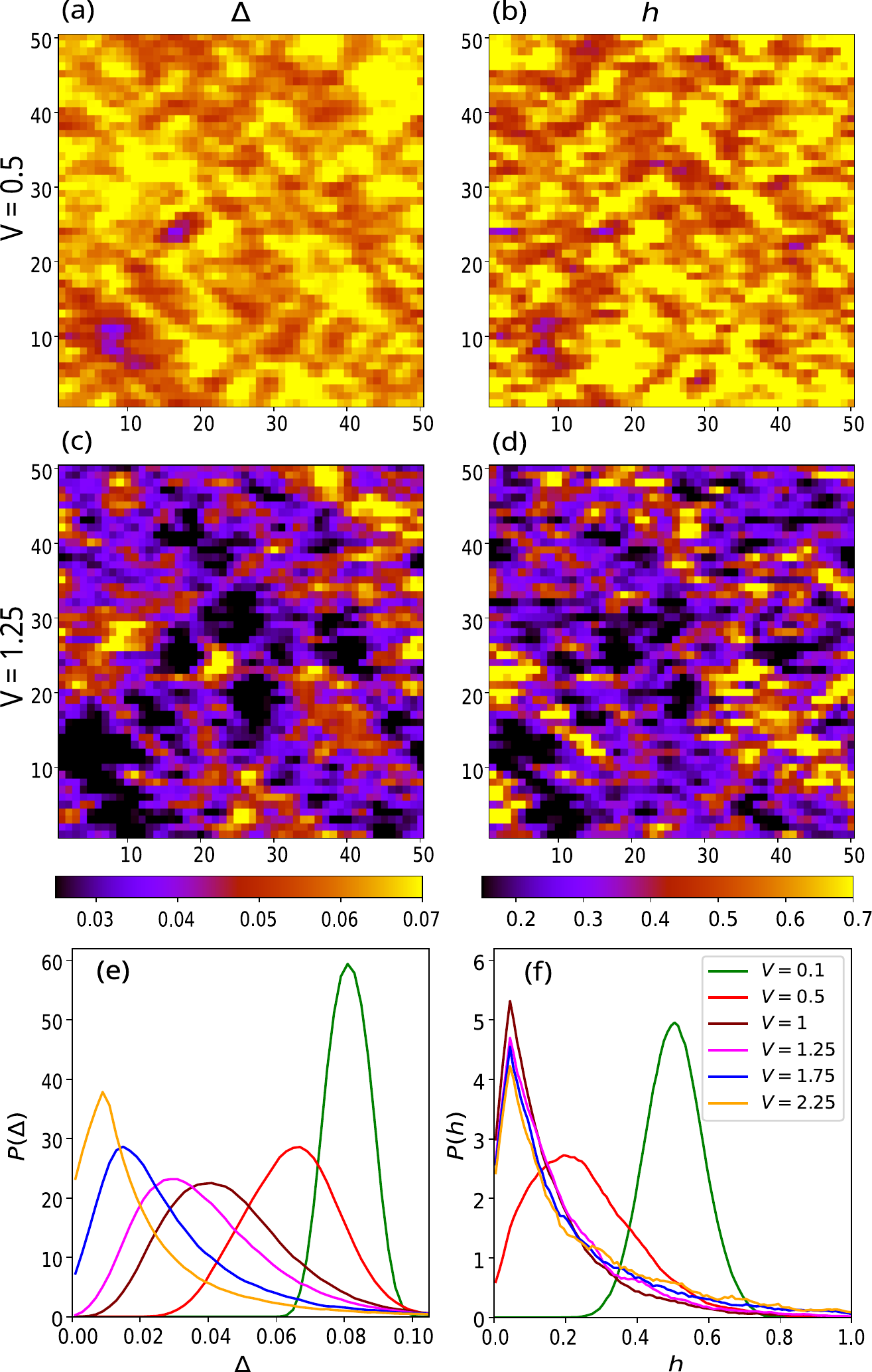}
    \caption{The spatial correlation of SC pairing amplitude $\Delta_i$ in panel (a), and the coherence peak height $h_i$ in panel (b) are demonstrated at a weak disorder strength $V=0.5$. Similar results for stronger disorder ($V=1.25$) are also shown in panels (c) and (d). The magnetic field is turned off ($\phi=0$) for all the results in this figure. In panels (e) and (f), we plotted the frequency (probability) distributions of $\Delta_i$ and $h$, respectively, where both $P(\Delta)$ and $P(h)$ broaden with increasing $V$, shifting the mode of these distributions toward zero, producing skewed distributions for larger $V$'s.}
    \label{onlydis}
\end{figure}
 %%%%__________________________________________________________
 %%%%%%%%%%%%%%%%%%%%%______________________________________________
 \begin{figure}[h]
    \centering
    \includegraphics[width=0.48\textwidth,height=1.4\linewidth]{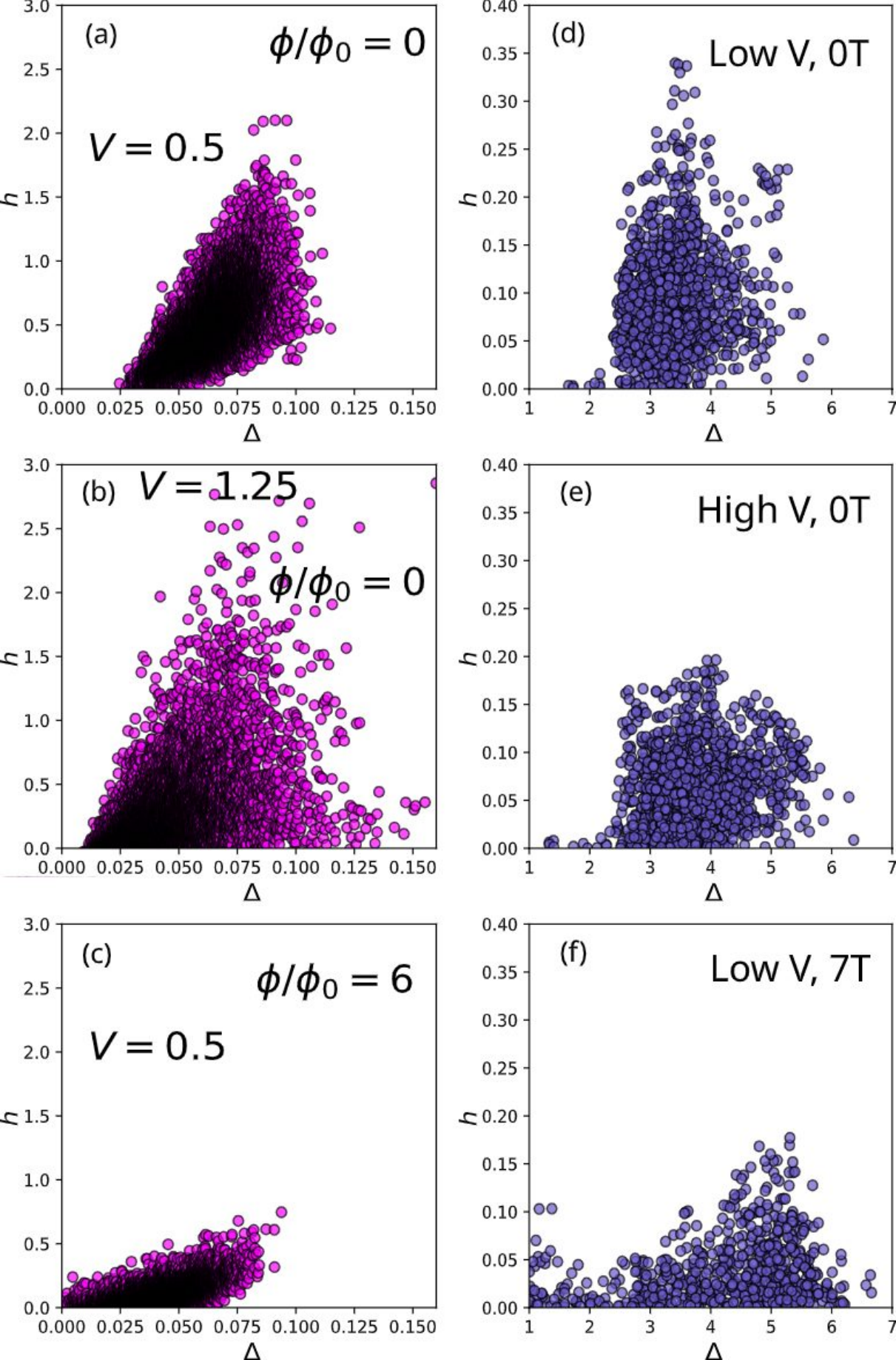}
    \caption{Scatter plots of the local order parameter $\Delta_i$, versus local coherence peak height $h_i$ with different $V$'s and $\phi$'s. Panels (a–c) present our theoretical predictions, while panels (d–f) show the results from our STS measurements. In both cases, the correlations between $\Delta_i$ and $h_i$ decrease systematically with increasing either $V$ or $\phi$, thereby establishing a close match between predictions and observations.    To quantify this observation, we calculated the Pearson correlation coefficient $r$ between $\Delta$ and $h$, and obtained $ r\sim 0.78$ for $V=0.5$ and $ r\sim 0.72$ for $V=1.25$ when $\phi=0$, and $ r\sim 0.66$ for $V=0.5$, $\phi/\phi_0=6$. Here, $r = \mathrm{cov}(\Delta, h)/\sigma_\Delta \sigma_h$ -- for details, see~\cite{note}. From experimental data, we find $r \approx 0.756$ at low $V$ and $0$ T, $r \approx 0.73$ at high $V$ and $0$ T, and $r \approx 0.70$ at low $V$ and $7$ T, which are in close agreement with the theoretical trends.}
    \label{scatter}
\end{figure}
%%%%_____________________________________________________________
We have determined the parameters $E_g^V$ and $C_V$ from the AvDoS for various values of $V$ and $\phi$. Because of the partial gap filling induced by $\phi$, resulting in a soft gap evidenced in Fig.~\ref{inhom-avdos}(b)[left], we also define a spatial profile of the LDoS at a sub-gap energy (which also reflects the inhomogeneity in the system), defined as
\begin{equation}
    N(i,E_g^V/2)=\frac{1}{2}\left(N(i,\omega=E_g^V/2)+N(i,\omega=-E_g^V/2)\right)
\end{equation}
For simplicity, we considered only one sub-gap energy scale, half of the corresponding energy gap.
\subsection{\label{subsec:Delta_h} Comparing inhomogeneities in SC pairing amplitude and coherence peak height}
As discussed, the SC pairing amplitude ($\Delta$), as well as CPH ($h$), become spatially inhomogeneous in response to the combined effects of $V$ and $\phi$. What are the similarities and dissimilarities in the inhomogeneous evolution of these observables as $V$ and $\phi$ are tuned? To address this question, we present our findings below on the spatial correlations of $\Delta_i$ and $h_i$.

\subsubsection{Evolution of spatial profiles of $\Delta$ and $h$ with $V$ for $\phi=0$}
We illustrated the spatial variation of $\Delta_i$ and $h_i$ for $V=0.5$ and $V=1.2t$ in Fig.~\ref{onlydis}(a, c), and Fig.~\ref{onlydis}(b, d) respectively. A comparison of panels (a) and (b) and similarly for (c) and (d) reveals a positive spatial correlation in them, i.e., a site with large $\Delta_i$ is more likely to have a large $h_i$, and vice versa, though such correlations are not exact. Also, such spatial correlations deteriorate further with increasing $V$. This trend is also evident in the scatter plots of ($\Delta_i,h_i$) for different $V$ as presented in the Fig.~\ref{scatter}(a-b). A comparable evolution of the $\Delta_i$–$h_i$ correlation with disorder has also been observed in the experiment, as shown in Fig.~\ref{scatter}(d,e).

The other measure of inhomogeneity is the probability (frequency) distribution of different inhomogeneous observables (similar to $P(\Delta)$ of Fig.~\ref{inhom-avdos}(a).
For example, in Fig.~\ref{onlydis}(e), we present $P(\Delta)$ for several $V$s. Such an evolution of $P(\Delta)$ with $V$ is well documented in the literature~\cite{PhysRevB.71.014514, PhysRevLett.81.3940}. Interestingly, we find that the frequency distribution of CPH, $P(h)$, also shows a qualitatively similar development with $V$ in Fig.~\ref{onlydis}(f). There are differences in the details, and such differences become more discernible at larger $V$. This is consistent with our previous discussions. 

\subsubsection{Evolution of spatial profiles of $\Delta$ and $h$ with $\phi$ at weak $V$}
\begin{figure}[h]
    \centering
    \includegraphics[width=0.5\textwidth,height=1.4\linewidth]{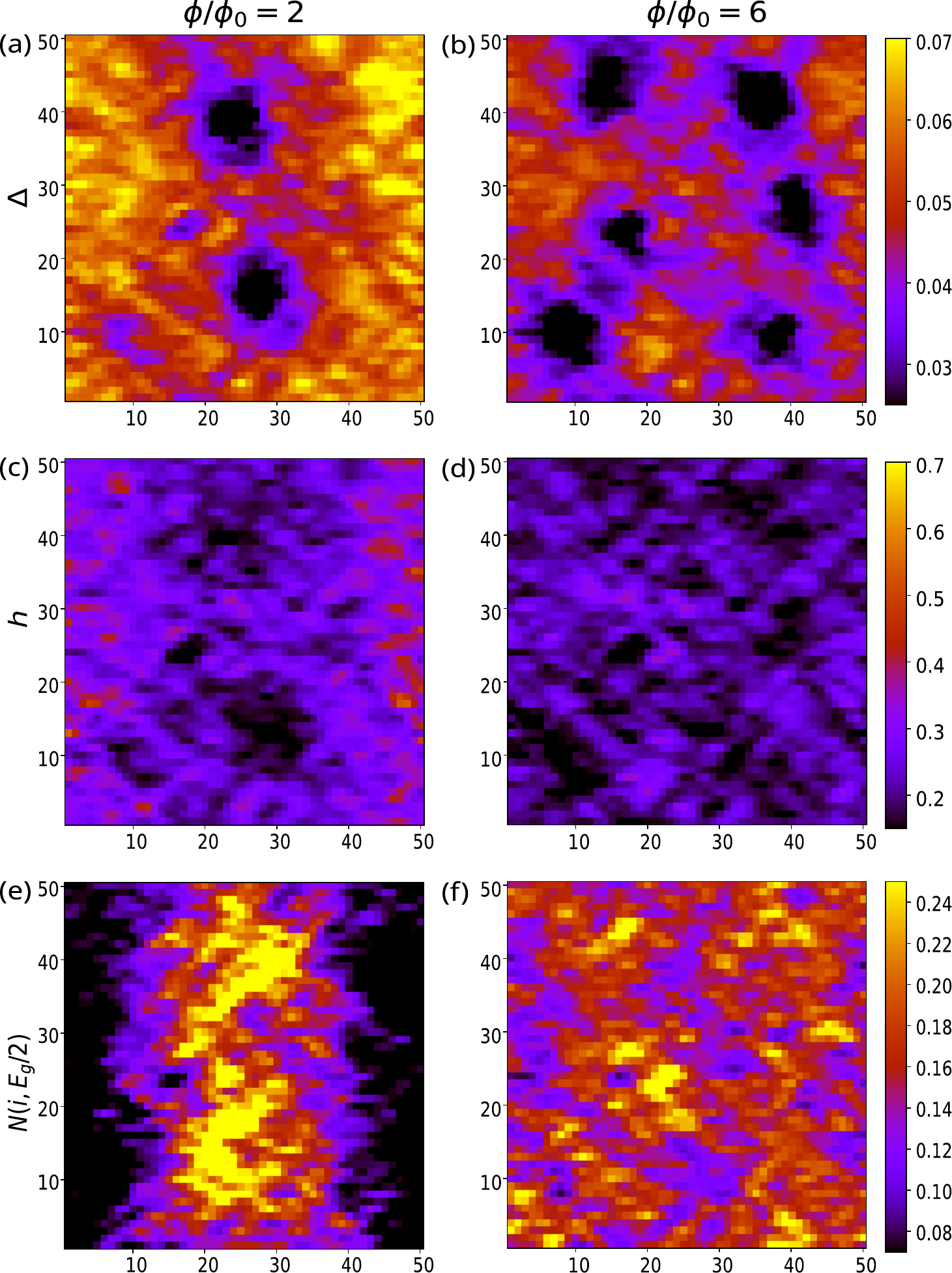}
    \caption{Panels (a,~c,~e) reflect the spatial map of $\Delta_i$, $h_i$, and  $N(r_i, E_g/2)$ respectively, for $\phi_/\phi_0 = 2$, whereas, the same observables are presented in panels (b,~d,~f) for a larger $\phi_/\phi_0 = 6$. All panels show results for $V=0.5$. In (a, b), we observe a significant suppression of SC pairing amplitude in the dark regions, due to the formation of two and six vortices, respectively. Outside the vortex cores, disorder induces weaker spatial fluctuations in $\Delta_i$. Similarly, in (c,d), the local coherence peaks are fully suppressed, while in (e,f), normal electronic states develop within the subgap region at and around the vortex cores.}
    \label{dismag}
\end{figure}

The $H$-dependence of the spatial correlations between $\Delta_i$ and $h_i$ at weak disorder ($V=0.5$) is depicted in Fig. \ref{dismag}(a-d). In addition, we include the behavior of $N(i, E_g/2)$ (Fig.~\ref{dismag}(e,f)), which is the local density of states evaluated at a \enquote{subgap} energy ($=E_g^V/2)$.

\begin{figure}[h]
    \centering
    \includegraphics[width=0.48\textwidth,height=1.4\linewidth]{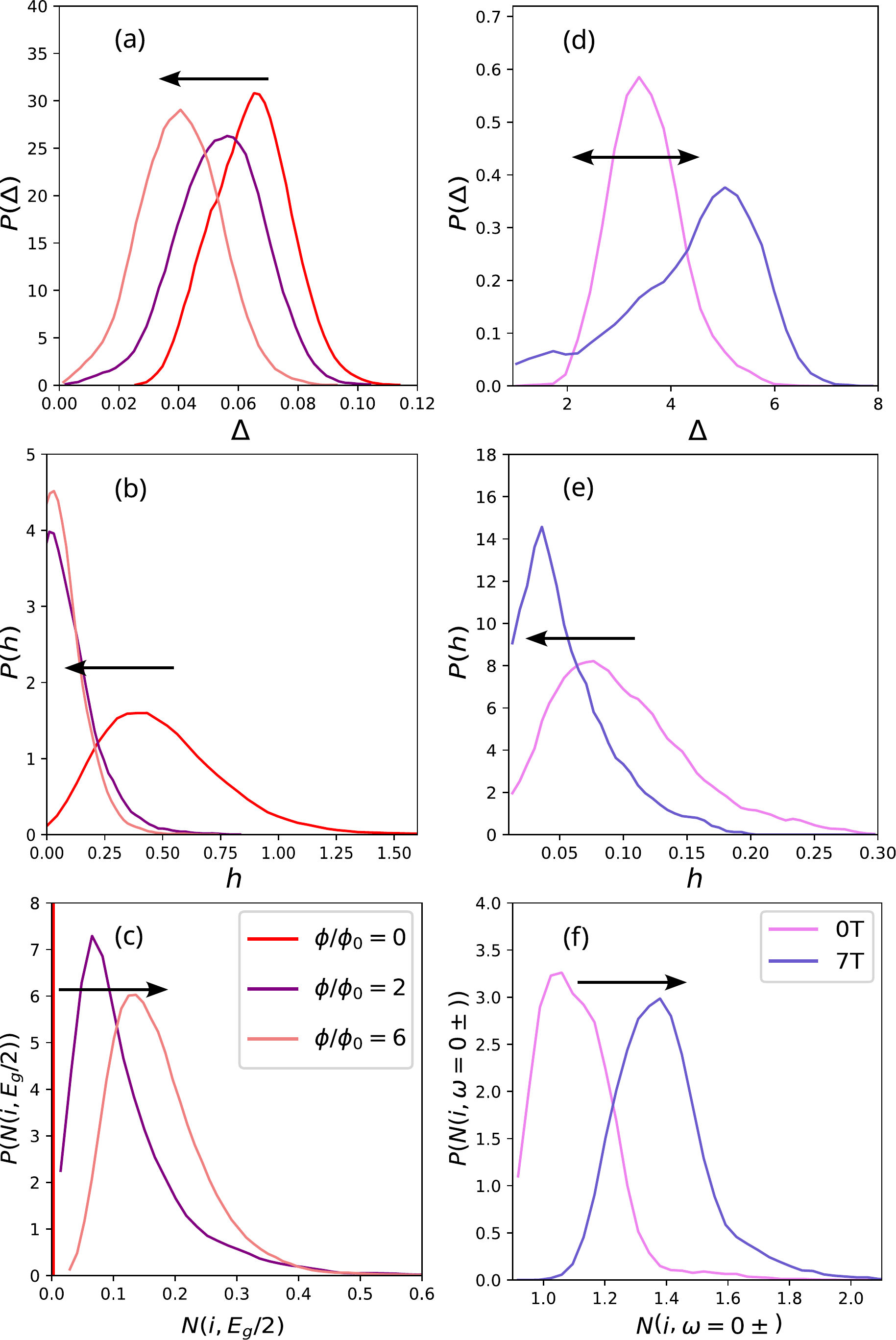}
    \caption{Distributions of local superconducting properties under varying magnetic fields. Panels (a–c) show the evolution of the probability distributions of the local SC pairing amplitude $\Delta_i$, the local coherence peak height $h_i$, and the subgap (actually, at $\omega=E_g/2$) LDoS $N(i, E_g/2)$, respectively, for our calculated estimates in the absence ($\phi/\phi_0=0$) and presence ($\phi/\phi_0=2,6$) of an orbital magnetic field for $V=0.5$. The distributions reveal enhanced spatial inhomogeneity and spectral broadening under field, reflecting vortex-induced suppression of coherence peaks and the emergence of low-energy states. Panels (d–f) show the corresponding distributions from the measured STS data obtained in the absence ($0$~T) and presence ($7$~T) of an orbital magnetic field.}
    \label{dismag_pdf}
\end{figure}

The scatter plot of spatial correlation between $\Delta_i$ and $h_i$ in Fig.~\ref{scatter}(c) shows that a positive correlation between the two becomes progressively weaker with increasing $\phi$. This is reflected in the spatial distribution of $\Delta_i$ and $h_i$ on the lattice (See Fig.\ref{dismag}). It is, however, reassuring that the predicted evolution from our simulation matches the experimental findings very well, as illustrated in Fig.~\ref{scatter}(f). The qualitative nature of $P(\Delta)$ remains unaltered as the magnetic field is turned on, as depicted in Fig.~\ref{dismag_pdf}(a). Still, the distribution approaches zero due to the suppression of the order parameter along field lines. Interestingly, this trend too is in good agreement with the experimental findings; see Fig.~\ref{dismag_pdf}(d). Turning to the distribution of the coherence peak height, $P(h)$: We find that the significant weight of the distribution shifts towards zero-value upon incorporating $\phi$ as shown in Fig.~\ref{dismag_pdf}(b). This result is also consistent with corresponding experimental findings [Fig.~\ref{dismag_pdf}(d)].

It is worth noting that We found a weak anti-correlation between $\Delta_i$ and $N(i, E_g/2)$, such that, regions with large $\Delta_i$ generally features a lower $N(i, E_g/2)$, while suppressed $\Delta_i$ (e.g., near vortex centers) corresponded to a higher $N(i, E_g/2)$. This indicates the formation of in-gap states resulting from the magnetic field. Our theory predicts a reshuffling of the weight of $P(N(i, E_g/2)$ upon introduction of $\phi$. In particular, the mode of this distribution shifts away from zero to a finite value. ($P(N(i, E_g/2, \phi/\phi_0=2))\approx 0.053$ and $P(N(i, E_g/2, \phi/\phi_0=6)\approx 0.13$); see Fig.~\ref{dismag_pdf}(c), stipulating the generation of low-lying states. 
%%%%%%%%%%%%%%%%%%%%%______________________________________________
%%%%%%%%%%%%%%%%%%%%%______________________________________________ 
The theoretical predictions also align well with the experimental observations, if we compare our numerical findings of $P(N(i, E_g/2))$ with $P(N(i,\omega=0\pm))$ obtained from experiments [Fig.~\ref{dismag_pdf}(f)]. However, our theory is bound to offer a near-zero weight for $P(N(i,\omega=0\pm))$.

Having emphasized the similarities between our theoretical predictions and experiments, we also acknowledge some dissimilarities. For example, unlike the experimental observations at zero field of low-lying states~\cite{article}, our BdG calculations with Hamiltonian ${\cal H}$ of Eq.~(\ref{dismag_pdf}), predict a hard gap in the AvDoS of SC films with non-magnetic impurities in the absence of $H$. As a result, $P(N(i, E_g/2))$ turns into a $\delta$-function near $N(i, E_g/2)=0$ (see Fig.~\ref{dismag_pdf}(c)). The experimental signature of considerable gap-filling (see Fig.~\ref{dismag_pdf}(f)) under similar conditions, which was argued to originate from paramagnetic impurities on the non-superconducting \ce{Sr2VO_{3-\text{x}}} layer of \ce{Sr2VO_{3-\text{x}}FeAs}.

Also, when $\phi$ is introduced, the centroid of the experimental distribution $P(\Delta)$ moves towards larger $\Delta$, relative to where it was at $\phi=0$ (see Fig.~\ref{dismag_pdf}(d)). This feature does not hold for our theoretical calculations [Fig.~\ref{dismag_pdf}(a)].
%______________________________________________
%______________________________________________________
\begin{figure*}[t]
    \centering
    \includegraphics[width=1\textwidth,height=0.65\linewidth]{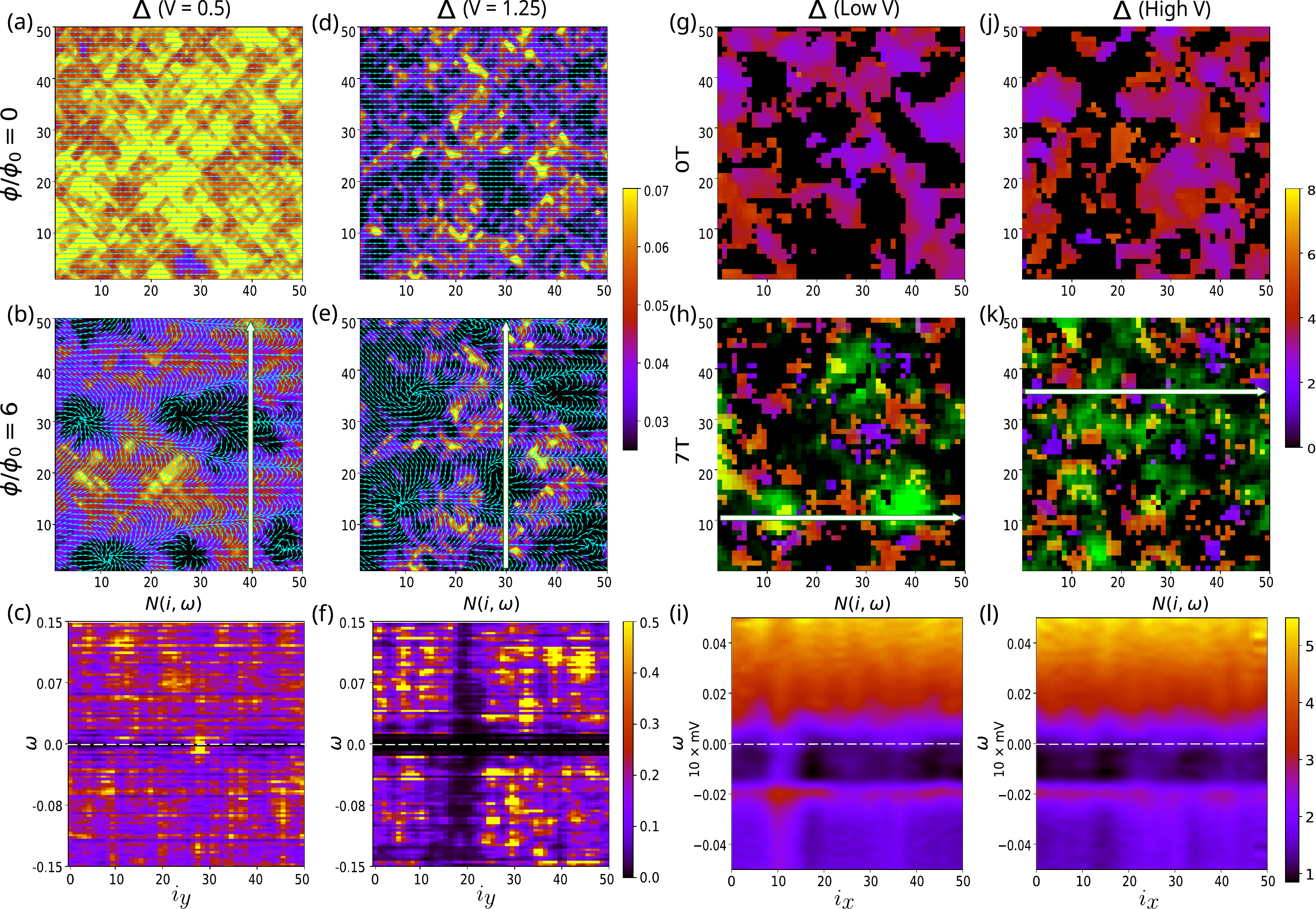}
    \caption{Panels (a) and (b) show the spatial profile of the SC order parameter ($\Delta_i$) for a single disorder realization ($V=0.5$) in the absence of $\phi$, and when $\phi/\phi_0=6$, respectively ( the pairing amplitude is shown as a color-density plot highlighting SC islands, which is overlaid with arrows to indicate the phase of the order parameter, identifying vortex positions). $\Delta_i$ exhibits a weakly inhomogeneous spatial profile in panel (a) due to small $V$, while in panel (b) dark regions indicate suppression of $\Delta$ due to vortex formation in this inhomogeneous background. In panel (c), we plot the LDoS for sites along the white vertical line ($x=39$) in panel (b). It exhibits a well-formed gap for sites along $i_y \in [0,25]$ and for $i_y \geq 32$, whereas subgap states appear for $26 \leq i_y \leq 31$. The vortex center is located at $(39,28)$ in panel (b). Panels (d–f) show similar results as in (a–c) but for stronger disorder $V=1.25$. In panel (f), the LDoS along the white line ($x=29$) in panel (e) exhibits behavior distinct from that in panel (c). Here, the LDoS exhibits a strong gap for $i_y \in [17,23]$ across the entire energy range, indicating that the flux line passes through a region of near-zero pairing amplitude, which is already suppressed due to strong $V$. Panels (g–k) present the local SC gap $\Delta$ extracted from LDoS, measured under low and high effective disorder conditions at $0$~T and $7$~T magnetic fields, demonstrating trends similar to the theoretical results. The overlaid green colors in panel (h,k) represent the height map of the zero-bias conductance peak from the experiment, indicating the position of vortex states. Panels (i) and (l) depict the LDoS along the white lines at $y=10$ in panel (h), and at $y=35$ in panel (k), respectively. In panel (i), accumulation of subgap states occurs for sites with $i_x \in [9,15]$ and $[32,40]$, corresponding to vortex formation. %(inferred from the LDoS data, see text).
    In panel (l), the vortex state appears in the range of sites with $i_x \in [26,32]$. In panels (i) and (l), the nearly symmetric peaks near $\pm 6$ meV are the SC coherence peaks.}
    \label{av-jv}
\end{figure*}
%______________________________________________________

\subsection{\label{subsec:av_jv} Abrikosov vs Josephson vortex in disordered SC}
While Abrikosov vortices (AV)~\cite{Abrikosov:1957wnz} are of common occurrence in bulk SCs in the presence of orbital magnetic fields, vortices of a different nature, known as Josephson vortices (JV), also arise, particularly in granular systems. A JV is characterized by quantized magnetic flux lines confined within Josephson junctions, characterized by a circulating supercurrent and a localized phase difference of the SC order parameter across the junction. It has been shown recently~\cite{PhysRevB.107.L140502} that in disorder SC, vortices transform from AV to JV as the disorder strength $V$ is increased. Here, we predict the spectroscopic signals associated with the $V$-induced transformation from an AV and JV from our BdG calculations, and compare such predictions with experimentally  obtained LDoS, accessed in STS analysis~\cite{article}.%}

Fig.~\ref{av-jv} illustrates the interplay of disorder and magnetic flux in shaping the local SC properties. For weak disorder ($V=0.5$), the SC pairing amplitude $\Delta_i$ is moderately inhomogeneous, and the LDoS exhibits a well-defined gap except near the vortex center, where subgap states emerge, consistent with the formation of conventional Abrikosov vortices.
%______________________________________________________
\begin{figure*}[t]
    \centering
    \includegraphics[width=1\textwidth,height=0.65\linewidth]{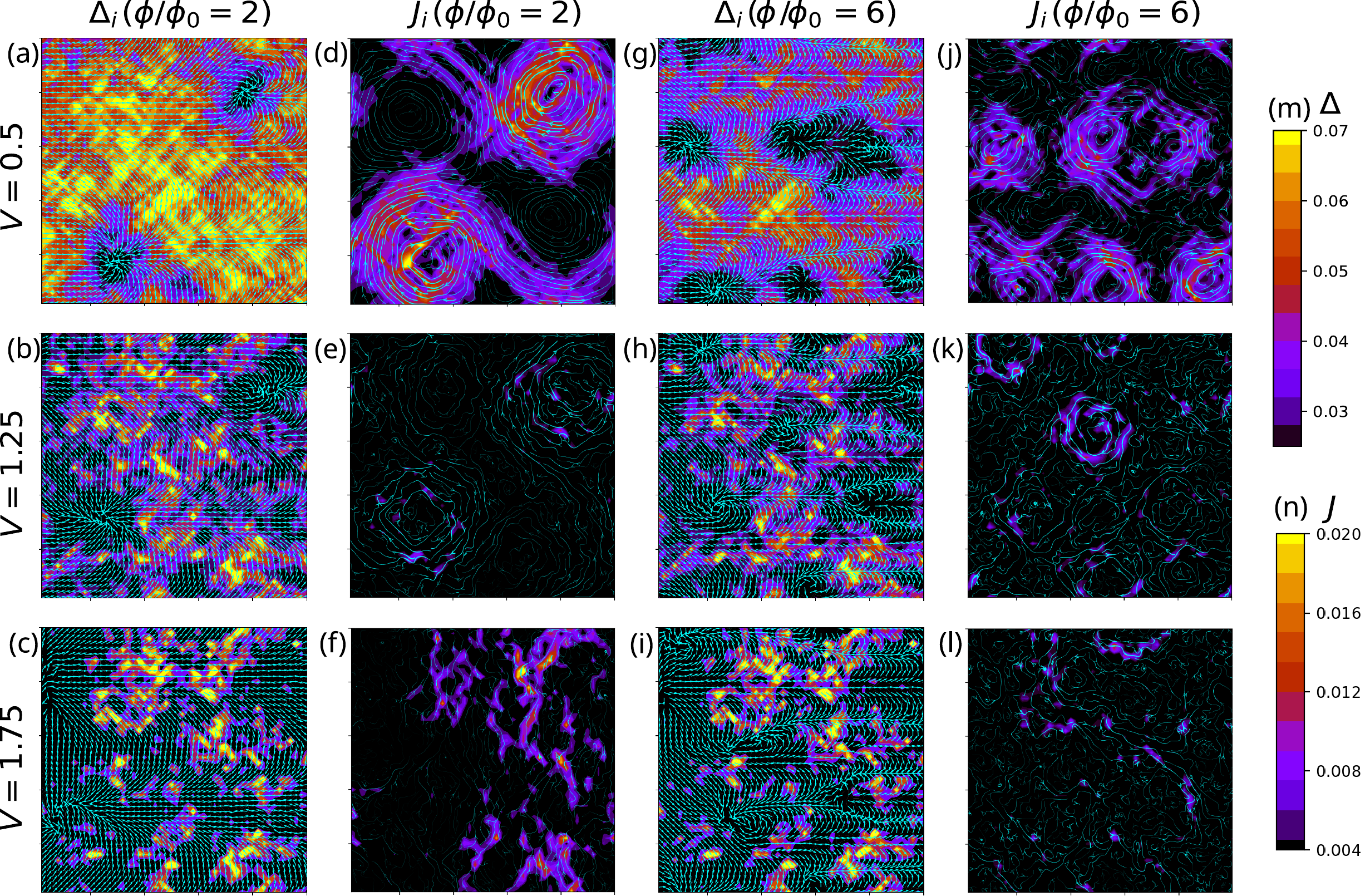}
    \caption{Panels (a-c) show the spatial profile of the SC pairing amplitude ($\Delta_i$), and (d-f) show the local supercurrent ($J_i$) for three disorder strengths ($V=0.5, 1.25, 1.75$) at $\phi/\phi_0=2$. Panels (g-i) and (j-l) present the corresponding profiles for $\phi/\phi_0=6$. Panels (m,n) provide the color-bar scales for $\Delta$ and $J$, respectively. In the $\Delta_i$ maps (1st and 3rd columns), the color-density plot refers to the magnitude of the order parameter, and the overlaid arrows indicate its phase. The $J_i$ maps (2nd and 4th columns) display current flow, where higher $\phi$ increases vortex density and alters current paths, while stronger disorder ($V$) induces vortex pinning and inhomogeneous, percolative transport. Arrows in $J_i$ plots indicate current direction, and green contours mark equal-magnitude regions with width proportional to supercurrent strength, highlighting vortices and flow inhomogeneity.}
    \label{current}
\end{figure*}
%______________________________________________________
In contrast, for strong disorder ($V=1.25$), the SC pairing amplitude is locally depleted even in the absence of an external field, creating regions that can accommodate phase twists at a minimal energy cost. In such regions, the system supports \enquote*{core-less} Josephson vortices, where the LDoS remains gapped across a wide energy range despite the presence of vortices. Theoretical analysis indicates that vortices preferentially form in regions of suppressed $\Delta_i$, thereby minimizing the bending energy associated with spatial variations of the order parameter.

Experimental STS measurements confirm these qualitative trends. At low effective disorder, the LDoS maps (see Fig.~\ref{av-jv}(i)) reveal enhanced spectral weight near $\omega \approx 0$ at vortex cores with locally suppressed SC pairing amplitude (Fig.~\ref{av-jv}(h)), consistent with Abrikosov vortices. As disorder increases, spatially extended regions of suppressed SC pairing amplitude emerge. Vortices in these regions no longer exhibit well-defined cores, consistent with the formation of core-less Josephson vortices with partially suppressed $\Delta_i$, as demonstrated in Fig.~\ref{av-jv}(k). The LDoS in these regions exhibits a gap (see Fig.~\ref{av-jv}(l)), yet still accommodates phase twists, in agreement with theoretical predictions.

These observations reveal a clear transition from Abrikosov to Josephson vortex behavior with increasing disorder. Our theoretical analysis of the LDoS and SC pairing amplitude maps reproduces the experimental trends, demonstrating that regions of quasi-locally suppressed $\Delta_i$ with a wide LDoS gap correspond to core-less Josephson vortices, while sites with locally suppressed $\Delta_i$ and strong subgap states indicate Abrikosov vortices. This agreement enables us to predict the nature of vortices directly from the theoretical LDoS, providing a unified framework for understanding vortex behavior in disordered SC under varying magnetic fields and disorder strengths.

\subsection{\label{subsec:currentdist} Evolution of supercurrent distribution under the combined influence of \textbf{V} and \textbf{H}}
In the clean limit, SC vortices yield local circulating currents induced by the applied magnetic field. These current loops around adjacent vortices cancel each other in equilibrium. Within our framework, the local supercurrent at site $i$ is defined as
\begin{equation}
    J_i={\rm Im}\left[t\sum_{j(\ne i),\sigma}(e^{i\phi_{ij}}\hat{c}^\dagger_{j\sigma}\hat{c}_{i\sigma}-e^{-i\phi_{ij}}\hat{c}^\dagger_{i\sigma}\hat{c}_{j\sigma})\right]
\end{equation}
which captures the net flow of Cooper pairs arising from phase gradients of the SC order parameter.

For small disorder ($V$) and weak magnetic field ($H$), the SC currents in our system remain largely coherent over long distances. However, the vortex lattice is no longer perfectly ordered due to the presence of disorder. The local current distribution (see Fig.~\ref{current}(d)) closely follows the smooth spatial variations of the SC pairing amplitude, $\Delta_i$ (shown in Fig.~\ref{current}(a)), reflecting a near-clean regime where phase coherence is mostly preserved despite mild spatial irregularities.

With increasing disorder, $\Delta_i$ develops strong spatial fluctuations (shown in Fig.~\ref{current}(b)), fragmenting it into isolated SC islands separated by non-SC domains. The resulting current flow (Fig.~\ref{current}(e)) becomes strongly inhomogeneous and follows percolative pathways through the SC regions.

As $H$ increases, vortices nucleate preferentially in regions where $\Delta_i$ is suppressed. However, in the presence of $V$, vortices get pinned increasingly irregularly (Fig.~\ref{current}(g-h)), distorting the standard supercurrent patterns of clean systems (Fig.~\ref{current}(j-k)), consistent with the irregular array of the flux lines.

At large $V$, JVs emerge, as discussed already, and correspondingly, the current dies out from most parts of the system. As seen from Fig.~\ref{current}(c), currents are dictated primarily by phase gradients rather than discrete vortex cores (Fig.~\ref{current}(f)). Here, currents remain limited to regions of significant SC pairing amplitudes.
When $V$ and $H$ are both strong, the current gets suppressed strongly due to the combined effects of overlap of vortex cores and Cooper pair localization, with $\Delta_i$ strongly diminished (see Fig.~\ref{current}(i)). This interplay eventually drives the system toward a superconducting-to-insulating transition, where global phase coherence is lost and supercurrent pathways are completely disrupted (Fig.~\ref{current}(l)).

% --- NEW 2–3 SENTENCES MOTIVATING FUTURE EXPERIMENTS (FROM ITEM 1) ---
Can these predictions of current paths upon tuning $V$ and $H$ be experimentally detected? Direct, nanometric visualization of these calculated current paths remains an open experimental goal. Josephson STM (JSTM, with a SC tip)~\cite{PhysRevB.93.161115} can map local Josephson coupling and phase-sensitive contrasts at atomic scales, but it does not image the supercurrent flow itself. Conversely, field-based probes such as scanning SQUID-on-tip ~\cite{Embon2015}, scanning NV magnetometry ~\cite{Rondin2013}, scanning Hall probes ~\cite{Ge2015}, and magnetic-force microscopy ~\cite{PhysRevB.98.140506} can recover current patterns but only indirectly via the stray-field (inverse Biot–Savart) and are typically limited to spatial resolutions of a few$\times10$~nm by sensor size and lift height. A combined approach, in which JSTM pins down phase stiffness while field-based Scanning Probe Microscopy (SPM) tracks the disorder- and field-tuned rearrangement of screening currents, would decisively test our predicted evolution from AV-like to JV-like current networks.
%%%%%%%%%%%%%%%%%%%%%%%%%%%%%%%%%%%%%%%%%%%%%%%%%%%%%%
\section{\label{sec:level4}Conclusion}
We presented a theoretical description of the evolution of inhomogeneities in SC correlations in a 2D film subjected to the combined influence of disorder and an orbital magnetic field. This is supplemented by our extraction of similar inhomogeneities from experimental STS data, wherever possible. A very satisfactory match between theoretical and experimental evolution of inhomogeneities gives us confidence in our comprehension of the underlying physics. We hope that our present
study will motivate the prediction of inhomogeneities in one observable, knowing it in another. Our findings elucidate the footprints of the disorder-driven transformation of the Abrikosov vortex to the Josephson vortex. In particular, we conclude that the signatures from LDoS and current patterns can conclude such a change in the nature of vortices quite extensively. 

%%%%%%%%%%%%%%%%%%%%%%%%%%%%%%%%%%%%%%%%%%%%%%%%%%
\section{ACKNOWLEDGMENT}
PS acknowledges support for the Ph.D. fellowship from the Council of Scientific and Industrial Research (CSIR), India. AG acknowledges support from the Indo-French Centre for the Promotion
of Advanced Research (CEFIPRA), Grant No. 6704-3. Numerical simulations were carried out using CQM (CEFIPRA grant) and KEPLER computing facilities at the Department of Physical Sciences, IISER Kolkata. JL and HRP acknowledge support by the Institute for Basic Science (Grant No. IBS-R014-D1). 

\bibliography{references}

\end{document}